\documentclass[twocolumn]{aastex63}
\usepackage{comment}
\newcommand{\zabs}{$z_{\rm abs}$}

\newcommand{\zem}{$z_{\rm em}$}

\newcommand{\kms}{km\,s$^{-1}$}
\newcommand{\cmsq}{cm$^{-2}$}

\newcommand{\Msun}{$M_{\odot}$} 
\newcommand{\ergps}{erg\,s$^{-1}$} 
\newcommand{\hi}{\mbox{H\,{\sc i}}}
\newcommand{\civ}{\mbox{C\,{\sc iv}}}
\def\lya{\ensuremath{{\rm Ly}\alpha}}

\submitjournal{ApJL}
\shorttitle{Powerful FRI-type quasar embedded in a \hi\ halo} 
\shortauthors{Gupta et al.}
\begin{document}


\title{\hi\ gas playing hide-and-seek around a powerful FRI-type quasar at $z\sim2.1$}

\correspondingauthor{N. Gupta}
\email{ngupta@iucaa.in}

\author{N. Gupta}  
\affil{Inter-University Centre for Astronomy and Astrophysics, Post Bag 4, Ganeshkhind, Pune 411 007, India}

\author{R. Srianand} 
\affil{Inter-University Centre for Astronomy and Astrophysics, Post Bag 4, Ganeshkhind, Pune 411 007, India}

\author{E. Momjian}  
\affil{National Radio Astronomy Observatory, P.O. Box O, Socorro, NM 87801, USA}

\author{G. Shukla} 
\affil{Inter-University Centre for Astronomy and Astrophysics, Post Bag 4, Ganeshkhind, Pune 411 007, India}

\author{F. Combes}  
\affil{ Observatoire de Paris, Coll\`ege de France, PSL University, Sorbonne University, CNRS, LERMA, Paris, France}

\author{J.-K. Krogager}  
\affil{Universit{\'e} Lyon1, ENS de Lyon, CNRS, Centre de Recherche Astrophysique de Lyon UMR5574, 69230 Saint-Genis-
Laval, France}

\author{P. Noterdaeme}  
\affil{Institut d'Astrophysique de Paris, UMR 7095, CNRS-SU, 98bis bd Arago, 75014  Paris, France}
\affil{Franco-Chilean Laboratory for Astronomy, IRL 3386, CNRS and Universidad de Chile, Santiago, Chile}

\author{P. Petitjean}  
\affil{Institut d'Astrophysique de Paris, Sorbonne Universit\'e and CNRS, 98bis boulevard Arago, 75014 Paris, France}

\begin{abstract}
We present optical spectroscopic and milli-arcsecond scale radio continuum observations of the quasar M1540-1453 (\zem = 2.104$\pm$0.002) that shows associated \hi\ 21-cm absorption at \zabs = 2.1139. At sub-kpc scales, the powerful radio source with 1.4\,GHz luminosity of $5.9\times10^{27}$\,W\,Hz$^{-1}$ shows Fanaroff-Riley (FR) class I morphology caused by the interaction with dense gas within 70\,pc from the AGN. 
Interestingly, while there are indications for the presence of absorption from low-ionization species like Fe~{\sc ii}, Si~{\sc ii} and Si~{\sc iii} in the optical spectrum, the expected strong damped \lya\ absorption is not detected at the redshift of the \hi\ 21-cm absorber. In comparison to typical high-$z$ quasars, the \lya\ emission line is much narrower. 
The `ghostly' nature of the \hi\ \lya\ absorber partially covering the broad line region of extent 0.05\,pc and the detection of widespread \hi\ 21-cm absorption covering the diffuse radio source (extent $>$ 425\,pc) imply the presence of a large clumpy \hi\ halo -- which may have been blown by the jet-ISM interaction. 
Further observations are needed to confirm the `ghostly' nature of the \lya\ absorber, and obtain a better understanding of the role played by the jet-ISM interaction in shaping the radio morphology of this powerful AGN.
The study showcases how joint radio and optical analysis can shed light on gaseous environment and origin of radio morphology in AGN at high redshifts, when these are still the assembly sites of giant galaxies.
\end{abstract}


\section{Introduction} 
\label{sec:intro}  

Through the detection of extended \lya\ emission, it is now well established that nearly all high-$z$ ($z>2$) quasars are embedded in low-density warm gas (T$\sim10^4$\,K) halos extending over few tens to hundred kpc \citep[e.g.,][]{Borisova2016}. The spatially resolved distribution and kinematics of this gas exhibits signatures of either infalling or outflowing material, potentially providing evidence for cold accretion flows and active galactic nuclei (AGN) feedback processes \citep[e.g.,][]{Arrigoni18, Shukla21}. 

Interestingly, quasar absorption-line studies suggest a very low covering factor of the high \hi\ column density ($N$(\hi)) gas around quasars.  Specifically, only 2-5\% of the quasars show a damped \lya\ system (DLA; with log~$N$(\hi)$\ge10^{20.3}$\,\cmsq) within $\pm$3000 \kms\ to the systemic redshift, \zem \citep[referred to as proximate DLAs or PDLAs;][]{Ellison02, Gupta21hz}. 
PDLAs associated with AGN may represent radiation feedback driven
gas outflows, and subsequent inflows i.e., `fountain' within the ionization cone shaped by the torus \citep[e.g.,][]{Wada12}. 
The leaked radiation from the broad- and/or narrow emission line regions (i.e., BLR and/or NLR) of the quasars can fill the \lya\ absorption trough and influence the detectability of such absorbers \citep{Finley2013}.
In the case of PDLAs identified through the presence of associated strong metal and H$_2$ absorption lines, 40-50\%
show detectable \lya\ emission in the DLA trough, implying that the absorbing cloud only partially covers the BLR \citep[e.g.,][]{Fathivavsari2018, Noterdaeme2019}. In certain cases, hereafter referred as `ghostly' DLA,  the PDLA close to the central engine is so compact ($<$1\,pc) that the leaked emission completely fills the \lya\ trough and its presence is confirmed only by the detection of the Ly$\beta$ absorption line \citep[e.g.,][]{Fathivasari17}.

Despite rarity, a complete census and detailed investigations of these absorbers provide crucial insights into different feedback processes operating in quasar host galaxies \citep[e.g.,][]{Noterdaeme2021b, Noterdaeme2021}. By revealing the chemical composition, clumpiness, and physical conditions in denser and colder phases of gas at parsec to kpc scales from the nucleus, these absorption line studies can bridge the gap in our understanding of how the ubiquitous halo gas and circumnuclear disk fuel the AGN activity \citep[see][for examples of nearby AGN where central regions are spatially resolvable in emission lines]{Combes14, Garcia-Burillo19, Storchi-Bergmann19} in distant quasars.

In the radio-loud regime, the \hi\ 21-cm line absorption toward parsec scale radio jet and extended lobe emission can complement the inherently one-dimensional exploration of optical/ultraviolet absorption line studies \citep[e.g.,][]{Adams09, Srianand15}.
Further, when the radio source is young i.e., still embedded within the host galaxy's interstellar medium (ISM), the properties of the ambient medium can also be used to understand the origin of Fanraoff-Riley Class I (FRI; edge darkened) and Fanaroff-Riley class II (FRII; edge brightened) division and evolution of radio jets \citep[e.g.,][]{Sutherland07, Wagner12}. The more powerful FRII sources typically have 1.4\,GHz spectral luminosity, $L_{\rm 1.4GHz}>10^{24}$\,W\,Hz$^{-1}$, while the FRI sources predominantly have lower luminosities. 
The FRI/FRII dichotomy is among the most debated issues in astrophysics, and it is unclear whether it originates from fundamental differences in the central engine i.e., black hole spin and accretion mode \citep[][]{Celotti97, Ghisellini01}, or jet composition \citep[][]{Reynolds96}, or jet deceleration caused by the ambient clumpy medium \citep[e.g.,][]{Bicknell1995, Mukherjee16}.

Here, we present optical spectroscopy and 1.5\,GHz milliarcsecond (mas)-scale imaging of a powerful radio quasar - M1540-1453 (\zem = 2.1; RA: 15:40:15.2353$\pm$0.0003; Dec: -14:53:42.409$\pm$0.011).  The radio source was identified as a high-$z$ quasar through a large optical survey of mid-infrared color-selected powerful radio sources \citep[][]{Gupta21salt}. The associated radio continuum emission corresponds to a 1.4\,GHz spectral luminosity, $L_{\rm 1.4GHz}$ = $5.9\times10^{27}$\,W\,Hz$^{-1}$ and radio loudness, $R$ = $f_{\rm 5GHz}/f_{2500\AA}$ = 4864). The associated black hole mass ($M_{BH}$) and Eddington ratio ($\lambda_{Edd}$) are 9.2$\times10^{8}$\,\Msun\ and 0.08, respectively \citep[][]{Gupta21salt}.

Besides being amongst the most powerful radio loud quasars (RLQs) known in the Universe \citep[][]{Gupta21salt}, M1540-1453 further stands out for the presence of a cold atomic gas reservoir revealed through the strong \hi\ 21-cm absorption \cite[$N$(\hi) = (2.06$\pm$0.01)$\times10^{21}$($\frac{T_{\rm s}}{100}$)($\frac{1}{f_{\rm c}}$)\,\cmsq\ at $z_{\rm 21cm}$ = 2.1139;][]{Gupta21hz}\footnote{Here $T_{\rm s}$ is the spin temperature in Kelvin and $f_c$ is the covering factor of the \hi\ gas.}.  The 90\% of the total \hi\ 21-cm absorption optical depth is within, $\Delta V_{90}$ = 167\,\kms.
At $z>2$, associated \hi\ 21-cm absorption has been detected in only three other cases \citep[][]{Briggs93, Moore99, Aditya21}.
In general, and based on sensitive low-frequency radio spectroscopic surveys with existing facilities, the powerful quasars at $2<z<3.5$ show an extremely small \hi\ 21-cm absorption detection rate \citep[$1.6_{-1.4}^{+3.8}$\%;][]{Gupta21hz}.
The joint radio and optical investigation of M1540-1453 presented here suggests that unlike the majority of powerful RLQs the radio jets in this case exhibit FRI morphology, are interacting with the host galaxy ISM, and are embedded in a \hi\ halo with signatures of  infall.  
Further, the absence of strong \hi\ absorption in the optical spectrum implies a possible `ghostly' DLA. 
%

Throughout this paper we use the $\Lambda$CDM cosmology with $\Omega_m$=0.315, $\Omega_\Lambda$=0.685 and H$_{\rm 0}$=67.4\,\kms\,Mpc$^{-1}$ \citep[][]{Planck20}. At $z = 2.1$, $1^{\prime\prime}$ = 8.5\,kpc.

\section{Observations and data analysis}      
\label{sec:obs}   

\subsection{Optical Observations}
\label{sec:optobs} 

\begin{figure*} 
\centerline{\vbox{
\centerline{\hbox{ 
\includegraphics[viewport=16 27 785 590, clip=true,width=0.95\textwidth,angle=0]{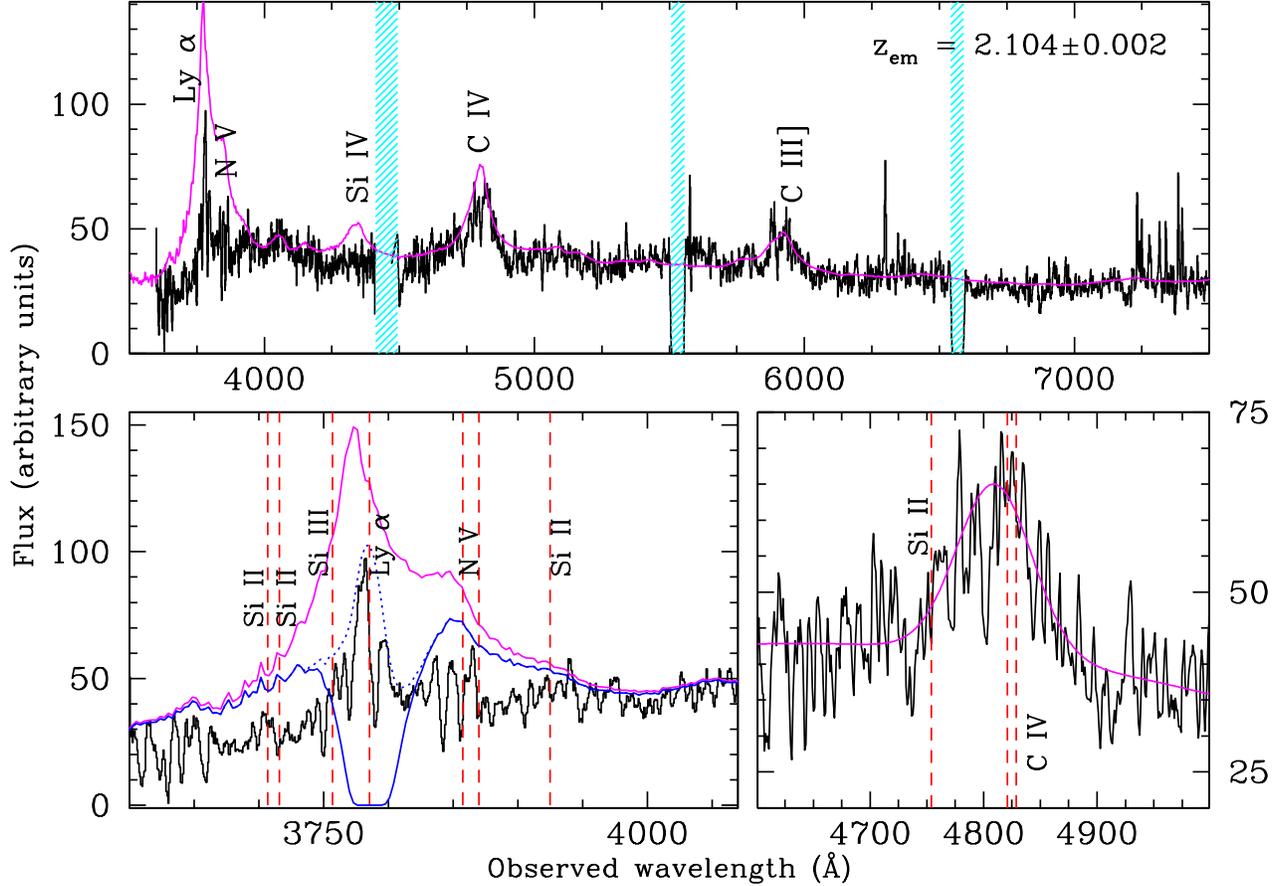} 
}} 
}}  
\vskip+0.0cm  
\caption{{\bf Top:} SALT spectrum of M1540-1453 overlaid with the SDSS template spectrum (magenta) of quasars tilted by $\lambda^{0.6}$. The locations of various emission lines are marked and cyan hashed regions correspond to the CCD gaps. It is evident that while the profiles of \civ\ and C~{\sc iii}] emission lines roughly follow the template, the \lya\ emission is relatively weak and narrow. {\bf Bottom left:} Spectrum around the the \lya\ emission with the vertical dashed lines show the expected locations of different low ionization absorption from the H~{\sc i} 21-cm absorber at \zabs = 2.1139. Solid blue curve is the predicted spectrum if the \hi\ 21-cm absorber (with $N$(\hi) = $10^{21.3}$\,\cmsq) completely covers the quasar emission.   {\bf Bottom right:} Gaussian fit (magenta curve) to the \civ\ profile after masking the absorption like features.
} 
\label{fig:salt_spec}   
\end{figure*} 

The optical discovery spectrum of M1540-1453 was obtained on 2015 August 07 using the Robert Stobie Spectrograph (RSS) on the Southern African Large Telescope (SALT) as part of a large survey of radio-bright ($>$200\,mJy at 1.4\,GHz) AGN \citep[details in][]{Krogager18, Gupta21salt} for the MeerKAT Absorption Line Survey \citep[MALS;][]{Gupta17mals}.  This spectrum covers 4486–7533\AA, that is, only C~{\sc iv}\,$\lambda$1549  and C~{\sc iii}]\,$\lambda$1909 emission lines.  The spectral resolution is $R=1064$ at 6041\,\AA. As the \civ\ emission line profile is affected by intrinsic absorption, we use a Gaussian fit to it to estimate the systemic redshift, \zem = $2.104\pm0.002$ (see bottom right panel in Fig.~\ref{fig:salt_spec}). This is lower than the redshift, \zem = 2.113, we quoted in \citep[][]{Gupta21hz} which was based on the peak of the \civ\ emission without correcting for the presence of absorption. Note that the peak of the C~{\sc iii}] emission line is also affected by absorption.

Intrigued by the detection of a strong \hi\ 21-cm absorption line, we obtained additional SALT spectra on 2021 April 4 and 7 covering the \hi\ \lya\ line (see Fig.~\ref{fig:salt_spec}). 
Interestingly, the \lya\ emission line from the quasar is  weak and narrow, and the expected strong DLA corresponding to the \hi\ 21-cm absorption is absent (see bottom left panel in Fig~\ref{fig:salt_spec}). 
However, as also mentioned in \citet{Gupta21hz}, we detect absorption coincidences at the expected wavelengths of Fe~{\sc ii} and Si~{\sc ii} (for the latter see bottom panels of Fig.~\ref{fig:salt_spec}).
We also detect associated \lya\ absorption coincident with \civ\ absorption lines at \zabs = 2.0994, 2.1027 and 2.1180. For \zabs =  2.1027, we also detect the N~{\sc v} doublets. 
For the above mentioned systemic redshift (\zem) based on the Gaussian fit to the \civ\ emission line these \zabs\ correspond to relative velocities in the range of $-$450 to $+1350$ \kms, and the \hi\ 21-cm absorption peak is redshifted by $+950$\,\kms.  The 1$\sigma$ uncertainty on \zem\ corresponds to a velocity shift of about $\pm$200\,\kms.  This implies that a significant portion of the gas detected in absorption is redshifted i.e., could be infalling with respect to the quasar.  This is subject to the caveat that  \civ\ emission line is known to underestimate the systemic redshift by several 100 \kms \citep[][]{Richards2011, Paris12}.

\subsection{Radio Observations}
\label{sec:optobs} 

\begin{figure} 
\centerline{\hbox{
\centerline{\vbox{ 
\includegraphics[
trim = {2.0cm 15cm 3.7cm 3.5cm}, clip=true,
width=0.50\textwidth,angle=0]{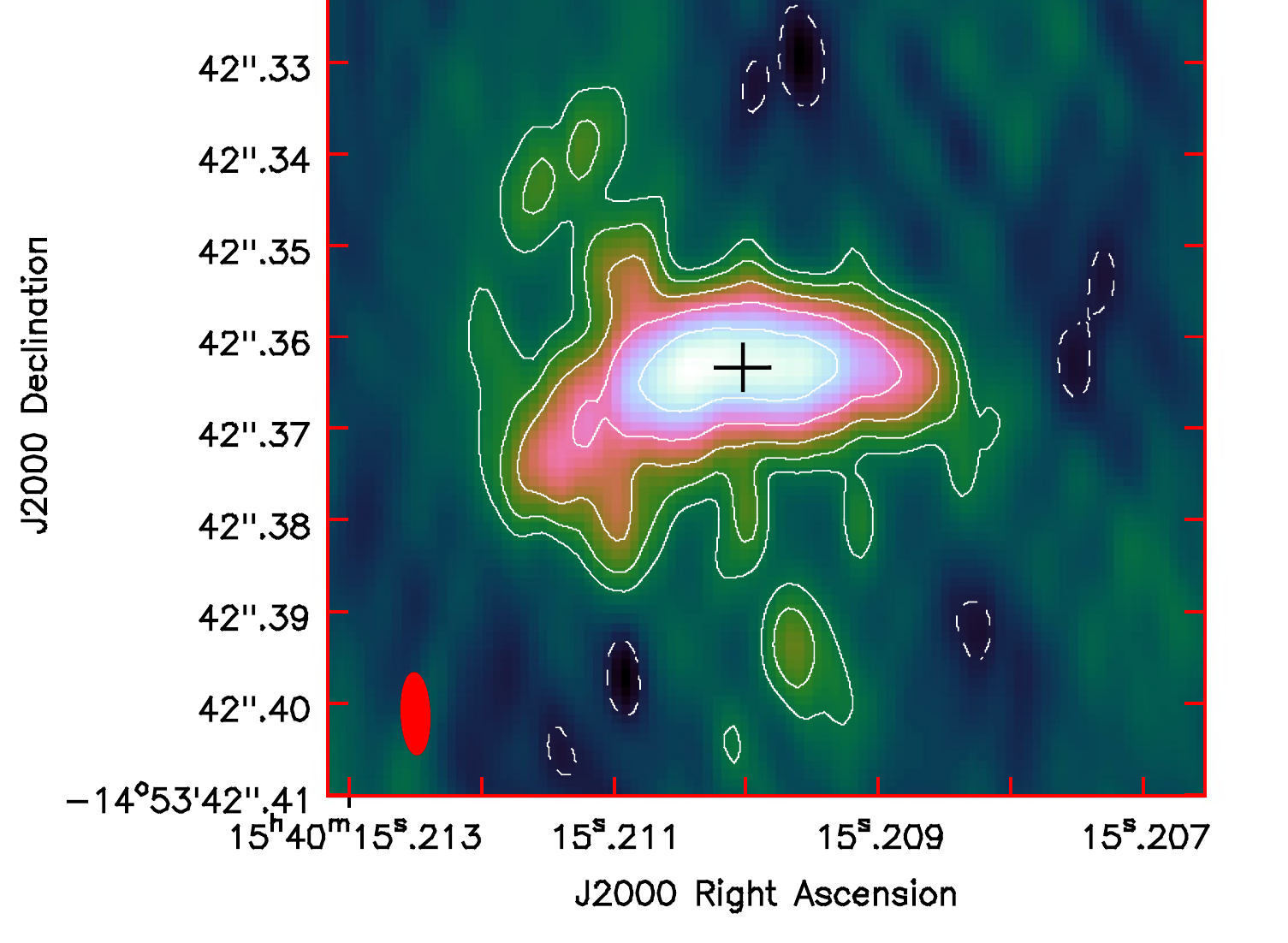}  
}} 
}}  
\vskip+0.0cm  
\caption{
VLBA 1.53\,GHz image of M1540-1453.  The rms noise in the image is 65\,$\mu$Jy\,beam$^{-1}$. The restoring beam of  9.0\,mas $\times 3.1$\,mas with a position angle of $+2.4^\circ$ is shown as an ellipse at the bottom left corner.  The contour levels are 250 $\times$ (-2, -1, 1, 2, 4, 8, 16, ...)\,$\mu$Jy\,beam$^{-1}$.  The peak intensity is 6.3\,mJy\,beam. `+' represents the location of putative radio core / AGN. 
} 
\label{fig:vlba}   
\end{figure} 

We observed M1540-1453 with the Very Long Baseline Array (VLBA) of the NRAO on 2021 June 9. A total bandwidth of 256\,MHz centered at 1.53\,MHz was used. At each of the 10 VLBA stations, eight 32\,MHz data channel pairs were recorded using the ROACH Digital Backend and the polyphase filterbank (PFB) digital signal-processing algorithm, in right- and
left-hand circular polarizations and sampled at two bits. The data were correlated with the VLBA DiFX correlator \citep[][]{Deller11} with 1\,s integration time. The observations employed nodding-style phase referencing using the nearby calibrator J1537-1527 with a cycle time of 3.67\,min. The source 3C\,345 was observed as a fringe-finder and bandpass calibrator. The total and on-source observing times were 2.0 and 1.3\,h, respectively. The data editing, calibration and imaging were done using the Astronomical Image Processing System (AIPS; \citealt{Greisen03}).  
The final image made using Briggs weighting with a robust factor of 0 in the AIPS task IMAGR is shown in Fig.~\ref{fig:vlba}.
The restoring beam is 9.0\,mas $\times 3.1$\,mas with a position angle of $+2.4^\circ$.
The total flux density measured with the VLBA is $\sim$75\,mJy and the extent of the detected radio emission is $\sim$50\,mas.

\section{Results}    
\label{sec:res}  

\subsection{Distorted radio morphology}    
\label{sec:radio}  

M1540-1453 has been detected as a compact radio source in a number of surveys from 100\,MHz to 3\,GHz. It has flux densities of 81.4$\pm$0.7\,mJy, 203.3$\pm$6.1\,mJy, 596$\pm$13\,mJy and 1172.9$\pm$13.0\,mJy at 3\,GHz \citep[Very Large Array Sky Survey (VLASS);][]{Lacy20}, 1.4\,GHz \citep[NRAO VLA Sky Survey (NVSS);][]{Condon98}, 420\,MHz \citep[upgraded Giant Metrewave Telescope (uGMRT);][]{Gupta21hz} and 170-231\,MHz \citep[GaLactic and Extragalactic All-sky MWA Survey (GLEAM) wide;][]{Hurley17}, respectively. Among these surveys, the VLASS provides the highest angular resolution. The Gaussian decomposition of the VLASS detected source implies a deconvolved size of $1.8^{\prime\prime}\times 0.7^{\prime\prime}$ (position angle = $155^\circ$). The spectral index between the VLASS and NVSS images is $\alpha^3_{1.4}$ = $-1.20\pm$0.05.  At lower frequencies ($<$1.4\,GHz), the GLEAM and uGMRT flux densities imply slight spectral flattening ($\alpha^{0.42}_{0.2}$ = $-0.91\pm$0.05). This may be an indication of a spectral turnover at much lower frequencies ($<$100\,MHz).

The radio source is resolved in our mas-scale resolution VLBA image at 1.53\,GHz (Fig.~\ref{fig:vlba}). At this frequency, only $\sim$41\% of the total flux density expected from the NVSS is recovered in the VLBA image. This implies that more diffuse emission, perhaps extending beyond 50\,mas ($\sim$425\,pc), is associated with the radio source. The extent of the projected radio emission, based on VLASS measurements, could be as large as 10\,kpc.  It is now well established that at kpc scales the FRII jets are better collimated than FRI jets \citep[e.g.,][]{Bridle1984}.  Both observations and hydrodynamic simulations suggest that radio jets in both  cases are initiated with relativistic speeds, but the decceleration caused by the entrainment of material disrupts the jet and may turn it from a FRII to FRI type morphology \citep[e.g.,][]{Parma1987, Bicknell1995}. 
%

In the case of M1540-1453, assuming the `+' in Fig.~\ref{fig:vlba} denotes the location of the radio core i.e., AGN, both the eastern and western jets are bright near the base and gradually fade into diffuse extended emission. No significant brightness asymmetry is observed on either side, but slight bends in the trajectory of jets at 4\,mJy\,beam$^{-1}$ (fifth contour in Fig.~\ref{fig:vlba}; $\sim$8\,mas i.e., $\sim$70\,pc from the core) are noticeable. These are reminiscent of the onset of dentist-drill effect beyond which the interaction with the ambient medium makes it harder for the jets to remain collimated and relativistic \citep[][]{Scheuer1974}.  Clearly, the radio jet has FRI morphology and is encountering substantial amount of gas at $\sim$70\,pc from the AGN. 
The similarities in jet morphology on either sides of the putative radio core suggest that the ambient media on both sides have similar properties.  

By definition, RLQs with highest radio powers are expected to be of FRII type \citep[e.g.,][]{Ghisellini01}. The FRI morphology of M1540-1453 is likely a direct consequence of the jet-ISM interaction.
The transition from FRII to FRI morphology due to jet-ISM interaction at sub-kpc scales have been noticed observationally in one case i.e., 3C\,84 \citep[e.g.,][]{Kino21} and in numerical simulations with flatter density gradients \citep[][]{Kaiser97, Wagner12}.

\subsection{Presence of a `ghostly' \hi\ absorber}    
\label{sec:abs}  

From the detection of strong \hi\ 21-cm absorption through uGMRT \citep[][]{Gupta21hz} and the VLBA morphology presented here, the AGN is expected to be buried in a gas rich environment. In the {\it top} panel of Fig.~\ref{fig:salt_spec}, we plot the scaled and reddened SDSS QSO template on top of the SALT spectrum.  It is evident from the figure that while the profiles of \civ\ and C~{\sc iii}] emission lines roughly follow the template, the \lya\ emission line associated with M1540-1453 is unusually narrow. 
In the {\it bottom left} panel of Fig.~\ref{fig:salt_spec}, we zoom in on the \lya\ emission and also plot the quasar spectrum modified by an \hi\ absorber with properties based on the 21-cm absorber (i.e., log\,$N$(\hi) = $10^{21.3}$\,\cmsq\ at $z_{21cm}$). Here, we have assumed $T_{\rm s}$ = 100\,K and $f_{\rm c}$ = 1.  Clearly, no such absorption is present along the optical sight line towards the AGN.  

However, as previously mentioned, the presence of low-ionization gas toward the optical sight line is confirmed through three Fe~{\sc ii} absorption features, and Si~{\sc ii} and Si~{\sc iii} absorption lines detected at $z_{21cm}$ (see Fig.~\ref{fig:salt_spec}).  This may indicate the presence of a `ghostly' strong \lya\ absorption associated with M1540-1453 \citep[][]{Fathivasari17}. 
The key requirement for the presence of a `ghostly' absorber is that the typical size of the absorbing cloud be smaller than the BLR but larger than the continuum emitting region at optical/UV wavelengths \citep[][]{Laloux21}.  For M1540-1453, we use $\lambda L_{1350}$ = $2.4\times 10^{45}$\,\ergps\ and Eq.\,2 of \citet[][]{Kaspi07} to estimate the BLR radius, $R_{\rm BLR}$ = 0.026\,pc.    
Thus, absorbing cloud towards the optical sight line ought to be smaller than $0.05$\,pc (diameter).  
The detection of damped Ly$\beta$ absorption at $z_{21cm}$ is needed to confirm the `ghostly' nature of this absorber.

\section{M1540-1453 and the ambient medium}    
\label{sec:amb}  

\begin{figure} 
\centerline{\hbox{
\centerline{\vbox{ 
\includegraphics[
trim = {3.5cm 6.0cm 6.0cm 5.0cm}, clip=true,
width=0.50\textwidth,angle=0]{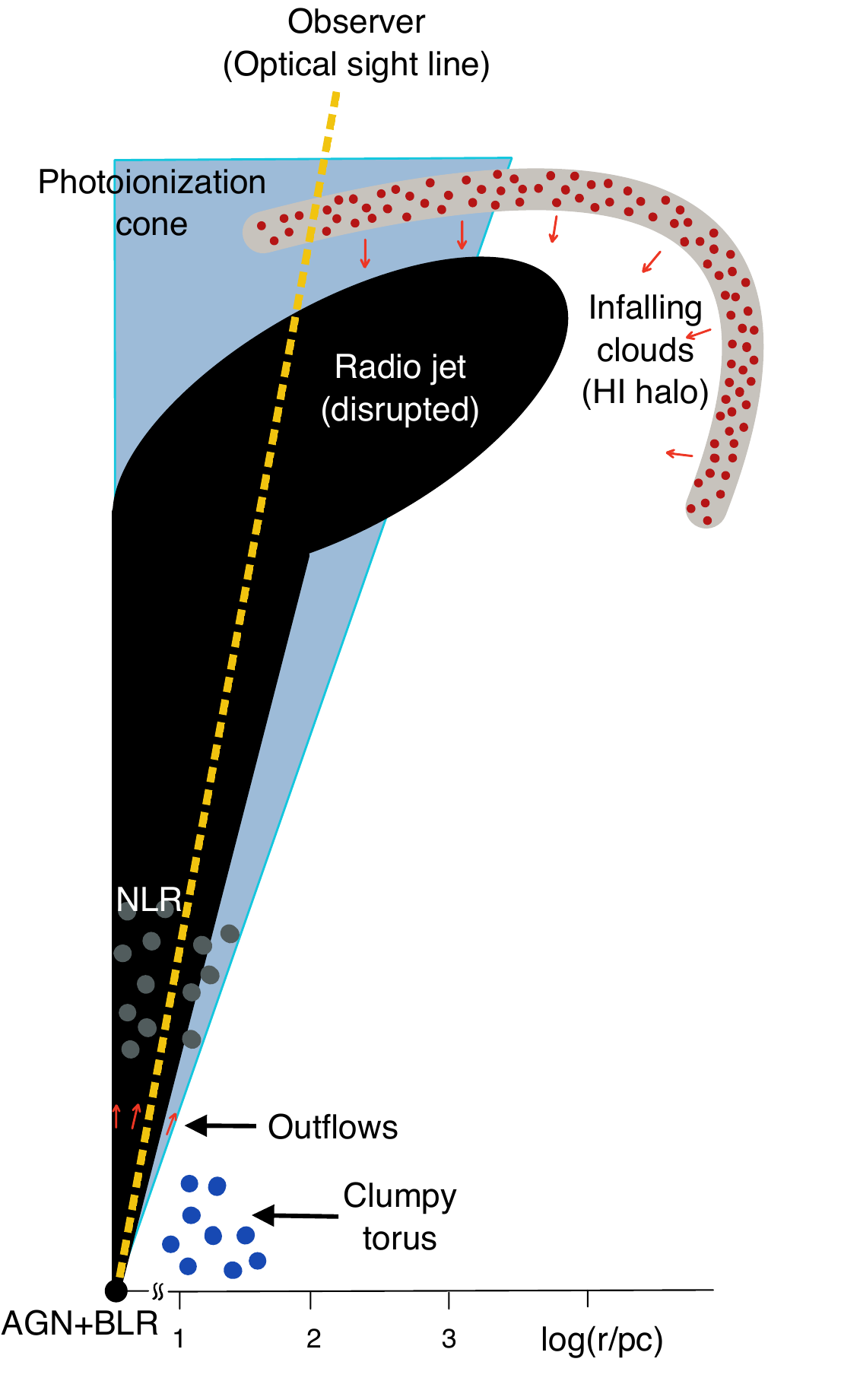} 
}} 
}}  
\vskip+0.0cm  
\caption{
Cartoon depicting a quadrant of M1540-1453 embedded in a clumpy ISM. The quasar's optical sight line (dashed line) may not intercept infalling high-$N$(\hi) clouds.  In comparison, extended radio emission ($>$425\,pc) from the lobes will be intercepted by several such clouds, producing the broad redshifted \hi\ 21-cm absorption line. The extent of \hi\ halo and the radio emission could be as large as 10\,kpc.
} 
\label{fig:cartoon}   
\end{figure} 
In general, the gas detected in \hi\ 21-cm absorption against a radio source could come from cold atomic gas in a regularly rotating structure i.e., circumnuclear disc or from the host galaxy ISM \citep[e.g.,][]{Carilli98m231}.  In the case of M1540-1453, the diffuse and distorted morphology of the radio emission (extent $>$425\,pc) seen in the 1.53\,GHz VLBA image, the broad ($\Delta V_{90}$=167 \kms) albeit smooth \hi\ 21-cm absorption profile and the lack of corresponding DLA in the optical spectrum imply the presence of numerous compact cold gas clouds forming a screen that covers the widespread radio emission and the AGN.  With the systemic redshift adopted in this work, the velocities of almost all the absorbing gas would be consistent with the infall scenario. It can be seen from Fig.~10 of \citet{Gupta21hz} that even if the \zem\ coincides with the peak of the \hi\ 21-cm absorption there is a significant redshifted wing corresponding to infalling cold \hi\ gas.

Using low-frequency flux density estimates from uGMRT and GLEAM, and the calibration of \citet[][]{Willott99}, we estimate the kinetic jet power associated with M1540-1453, P$_{jet}$ = $3\times10^{38}$ ($L_{\rm 151}/10^{28}$W\,Hz$^{-1}$) $\approx10^{46}$\,\ergps.  This is comparable to the radiative power represented by the bolometric luminosity \citep[$L_{\rm Bol}$ = $9.1\times10^{45}$\,\ergps;][]{Gupta21salt} of the AGN.  Thus, both the kinetic (jet-mode) and radiative (quasar-mode) feedback can affect the ambient gas here.  The radiative feedback alone can generate velocities of the order of -445 to +1350\,\kms\ observed in the optical spectrum \citep[e.g.,][]{Faucher-Giguere12}.

Further, the total energy deposited by the radio AGN over its lifetime ($10^{6-7}$\,yrs) through the interaction with the dense gas amounts to $10^{59-60}$\,erg.  In numerical simulations, this is adequate to drive a spherical energy-driven bubble of gas mass with observed velocities \citep[][]{Sutherland07}.  Following this phase the radio jet drives into the lower density ISM. The radiatively and kinetically accelerated gas clouds are expected to eventually fall back into the central regions over timescales of tens of Myr \citep[e.g.,][]{Wada12, Mukherjee16}.
The signatures of infalling gas in the present data are consistent with the above described scenario. Future near infrared spectroscopy and multi-frequency parsec-scale imaging of the radio continuum will provide an accurate systemic redshift measurement and constrain the orientation of the radio source and extent of the \hi\ halo, required to test this model for M1540-1453.

All the above considerations lead to a scenario summarized in Fig.~\ref{fig:cartoon}.  Radio jets are often observed to avoid extended circumnuclear discs but the jet-disc rotation axis may not be well aligned due to a variety of reasons.  Instead, the jet-axes are often found to be distributed over a wide range of angles within the polar cap  \citep[e.g.,][]{VerdoesKleijn05, Ruffa20}. In the simplistic case of M1540-1453 presented here, it is assumed that initially the jet is launched along the axis of the inner accretion disc.  However, at about $\sim$70\,pc from the `core' it gets misaligned due to interaction with the ambient medium.  This misalignment and the diffuse nature of radio emission has helped in probing cold gas over a large region, well beyond the initial jet axis.  Otherwise, this reservoir, and the presence of possible `ghostly' \hi\ absorber in the optical spectrum, would go unnoticed.

A similar scenario has been inferred for the well studied high-redshift radio galaxy (HzRG) B2\,0902+34 at $z\approx3.39$.  The associated radio source, FRII-type in this case, of $5^{\prime\prime}$ ($\sim$40\,kpc) extent has an orientation of $30^\circ$ to $45^\circ$ relative to the observer and is embedded in a huge \lya\ halo \citep[extent$\sim$100\,kpc;][]{Lilly88, Carilli95}. The detailed modelling of detected \lya\ halo and \hi\ 21-cm absorption in this case suggests that the radio AGN and the \lya\ halo are embedded in an even larger infalling neutral halo of \hi\ mass $>10^{12}$\,$M_\odot$  \citep[][]{Adams09}.  Crucial to the infall scenario are the shape of  \lya\ profile, inclination of radio source, and redshifts of \lya\ emission and \hi\ 21-cm absorption. Despite the relatively large inclination for an HzRG, the central AGN in the case of B2\,0902+34 is obscured suggesting that it is yet to blow out its environment. M1540-1453 shows no such obscuration implying that the AGN feedback has  already cleared the obscuring material and blown an \hi\ bubble, which depending on the accuracy of the systemic redshift, represents infalling material.  It is therefore caught at the later stages of evolution. It may also be the case that B2\,0902+34 is a protogiant elliptical galaxy in the early phases of evolution, and the collapsing structure represented by the \hi\ shell in this case has  an entirely different origin \citep[][]{Adams09}.

\section{Concluding remarks}    
\label{sec:conc}  

At low-$z$ ($z<0.3$), FRI-type radio sources are typically associated with low-excitation radio galaxies (LERGs) exhibiting low accretion rates ($\lambda_{Edd}<$0.01), low radio power ($L_{\rm 1.4GHz}<10^{25}$\,W\,Hz$^{-1}$) and hosted by red, massive early-type galaxies ($M_*\ge10^{11}$\,\Msun) \citep[][]{Best12}. M1540-1453, in comparison, is amongst the most powerful radio sources ($L_{\rm 1.4\,GHz}$ = 5.9$\times10^{27}$\,W\,Hz$^{-1}$)  and shows high-excitation optical spectra and efficiently accreting matter ($\lambda_{\rm Edd}$ = 0.08). Indeed, the investigation of \citet[][]{Best12} shows that at $L_{\rm 1.4\,GHz}$ = $10^{26}$\,W\,Hz$^{-1}$, LERGs may be as prominent as high-excitation radio galaxies (HERGs).  
M1540-1453 emphasizes jet-ISM interaction at sub-kpc scales as a pathway to produce FRI-type morphology, and may be more prevalent at high redshifts ($z\ge2$), when these are still the assembly sites of giant galaxies. 

Finally, we note \hi\ 21-cm absorption detections in AGNs similar to M1540-1453 are rare \citep[][]{Gupta21hz}.  Further work is needed to understand the curious case of this AGN presented here.  Does it represent a very short phase in the evolution of AGN, or simply a geometrical coincidence?  
The answers may lead to a better understanding of the sub-kpc scale environments through which FRI/FRII jets associated with powerful AGN are evolving at high-$z$.

\acknowledgments

We thank the anonymous referee for useful comments
and suggestions. This  work  is  based  on  observations  made  with  SALT and VLBA.  The National Radio Astronomy Observatory is a facility of the National Science Foundation operated under cooperative agreement by Associated Universities, Inc. \\

\facilities{SALT, VLBA}
\software{AIPS: \citep[][]{Greisen03} }


\def\aj{AJ}%
\def\actaa{Acta Astron.}%
\def\araa{ARA\&A}%
\def\apj{ApJ}%
\def\apjl{ApJ}%
\def\apjs{ApJS}%
\def\ao{Appl.~Opt.}%
\def\apss{Ap\&SS}%
\def\aap{A\&A}%
\def\aapr{A\&A~Rev.}%
\def\aaps{A\&AS}%
\def\azh{AZh}%
\def\baas{BAAS}%
\def\bac{Bull. astr. Inst. Czechosl.}%
\def\caa{Chinese Astron. Astrophys.}%
\def\cjaa{Chinese J. Astron. Astrophys.}%
\def\icarus{Icarus}%
\def\jcap{J. Cosmology Astropart. Phys.}%
\def\jrasc{JRASC}%
\def\mnras{MNRAS}%
\def\memras{MmRAS}%
\def\na{New A}%
\def\nar{New A Rev.}%
\def\pasa{PASA}%
\def\pra{Phys.~Rev.~A}%
\def\prb{Phys.~Rev.~B}%
\def\prc{Phys.~Rev.~C}%
\def\prd{Phys.~Rev.~D}%
\def\pre{Phys.~Rev.~E}%
\def\prl{Phys.~Rev.~Lett.}%
\def\pasp{PASP}%
\def\pasj{PASJ}%
\def\qjras{QJRAS}%
\def\rmxaa{Rev. Mexicana Astron. Astrofis.}%
\def\skytel{S\&T}%
\def\solphys{Sol.~Phys.}%
\def\sovast{Soviet~Ast.}%
\def\ssr{Space~Sci.~Rev.}%
\def\zap{ZAp}%
\def\nat{Nature}%
\def\iaucirc{IAU~Circ.}%
\def\aplett{Astrophys.~Lett.}%
\def\apspr{Astrophys.~Space~Phys.~Res.}%
\def\bain{Bull.~Astron.~Inst.~Netherlands}%
\def\fcp{Fund.~Cosmic~Phys.}%
\def\gca{Geochim.~Cosmochim.~Acta}%
\def\grl{Geophys.~Res.~Lett.}%
\def\jcp{J.~Chem.~Phys.}%
\def\jgr{J.~Geophys.~Res.}%
\def\jqsrt{J.~Quant.~Spec.~Radiat.~Transf.}%
\def\memsai{Mem.~Soc.~Astron.~Italiana}%
\def\nphysa{Nucl.~Phys.~A}%
\def\physrep{Phys.~Rep.}%
\def\physscr{Phys.~Scr}%
\def\planss{Planet.~Space~Sci.}%
\def\procspie{Proc.~SPIE}%
\let\astap=\aap
\let\apjlett=\apjl
\let\apjsupp=\apjs
\let\applopt=\ao
\bibliographystyle{aasjournal}
\bibliography{mybib}

\end{document}